# Non-equilibrium intermixing and phase transformation in severely deformed Al/Ni multilayers

X. Sauvage[1*], G.P. Dinda[2§], G. Wilde[2,3*]

1- Groupe de Physique des Matériaux - UMR CNRS 6634, Institute of Material Research, Université de Rouen, 76801 Saint-Etienne-du-Rouvray, France.
2- Institute of Nanotechnology, Forschungszentrum Karlsruhe, 76021 Karlsruhe, Germany
3- Institute of Materials Physics, University of Münster, 48149 Münster, Germany
§- *present address*: Department of Mechanical Engineering, University of Michigan, Ann Arbor, MI 48109–2092, USA

**Abstract**

Al/Ni multilayers have been prepared by repeated folding and cold rolling (F&R) of elemental foils. The thickness of Al and Ni foils is reduced down to less than 50 nm after fifty F&R cycles. Three-dimensional atom probe analyses clearly reveal the presence of supersaturated solid solutions and give the evidence of deformation-induced intermixing. The formation of the solid solutions and their transformation into the $Al_3Ni$ phase upon annealing is discussed.

\* corresponding authors:

xavier.sauvage@univ-rouen.fr          gwilde@uni-muenster.de

Tel. : + 33 2 32 95 51 42             + 49 251 83 33571

Fax. : + 33 2 32 95 50 32             + 49 251 83 38346

Phase transformations in thin metallic multilayer systems have been extensively investigated in the past. During heating of multiphase structures synthesized in thin multilayers, stable or metastable intermediate phases might appear as a result of interdiffusion and subsequent phase transformations. Aside from "static" microstructure features, interdiffusion or intermixing prior to phase transformation, and associated concentration gradients play a key role in determining the phase formation kinetics [1, 2]. It is well known that severe plastic deformation can induce significant mixing, as in ball milled powders [3-7]. Similar features have also been reported for torsion under high pressure [8] and cold rolling [9]. Since this latter technique is also suitable for the preparation of metallic multilayers [9-12], it was used to prepare Al/Ni multilayers to clarify the influence of concentration gradients and the intermixing state on the phase transformation kinetics upon annealing. The Al-Ni system was chosen as a model system since it is both attractive for applications, e.g. in superalloys or as coatings of III-V semiconductors, and has also been studied thoroughly concerning the microstructure– and the phase evolution [10, 11, 13-21]. In the studies reported so far, thin Al/Ni multilayers have mostly been deposited by sputtering [17, 19, 20] or electron-beam evaporation [16, 19-21], while repeated folding and cold rolling of Al and Ni foils was also applied as an alternative process to synthesize Al-Ni multilayer foils [10-12, 14, 18]. By this method, layers with thicknesses in the nanometer range were obtained and it was shown that such samples exhibit very similar phase transformation sequences and almost identical kinetic behaviour upon annealing compared to thin film multilayers that had been synthesized by deposition methods [12]. Yet, in the literature, different phase formation sequences have been reported for different nominal compositions and after different processing pathways. Under most conditions, the $Al_3Ni$ phase with an ordered $D0_{20}$ structure is the first phase to form. This is explained by the asymmetry of the interdiffusion coefficients, with Ni diffusing much more rapidly into Al than Al into Ni [22, 23]. However, depending on the nominal composition of the multilayer and on the processing history, also different initial intermediate phases such as AlNi [19, 20, 24-26], $AlNi_3$ [10] or a metastable $Al_9Ni_2$-phase [24, 27] were observed. In this context, it is interesting to analyze whether the phase formation sequence is determined by the processing route, and especially by the presence of concentration gradients prior to annealing. Three-dimensional atom probe (3D-AP) analyses were performed to map out these concentration gradients at the atomic scale on F&R-processed multilayers prior and after thermal annealing. In particular, we have investigated how local modifications of the distribution of constituents and the presence of steep concentration profiles can change the kinetics of intermediate phase formation.

Al/Ni multilayer samples of nominal composition $Al_{80}Ni_{20}$ were prepared from high purity foils of Ni (99.9%, thickness: 20μm) and Al (99.9%, thickness: 100μm) that were alternately stacked to form a 20 mm x 20 mm and 500 μm thick sandwich. This sample was then cold rolled (roll diameter 150 mm, 2 rotations per minute) to a thickness of 200μm in 10 passes, then folded to double the thickness and rolled again to a thickness of 200μm. This procedure was repeated 50 times. The final material was investigated in the as prepared state and after short time annealing in a differential scanning calorimeter (DSC, Pyris 1, Perkin Elmer) under Argon flow for 2 min at 250°C. This temperature was chosen because it is close to the onset for the $Al_3Ni$ formation [12, 14, 17, 18], and the duration was sufficiently short to avoid significant grain growth. Phase identification was carried out using X-ray diffraction (Phillips X'Pert) with a Cu anode source operating at 50kV and 40 mA. The microstructure of the multilayer was examined in a FEI/Tecnai F20 ST transmission electron microscope (TEM) operating at 200kV and an extraction voltage for the field-emission gun of 4.1 kV, using a Philips double-tilt holder. TEM samples were prepared in cross section geometry by ion milling using a Gatan Precision Ion Polishing System (PIPS 691) operating at 3.5 kV and an angle of incidence of 4°. 3D-AP analyses were performed with a CAMECA tomographic atom probe detection system (TAP) and analyses were carried out at 80 K with a pulse fraction of 16 % in UHV conditions. 3D-AP samples were prepared by electropolishing (20°C, 10V, 2% perchloric acid in 2-butoxyethanol).

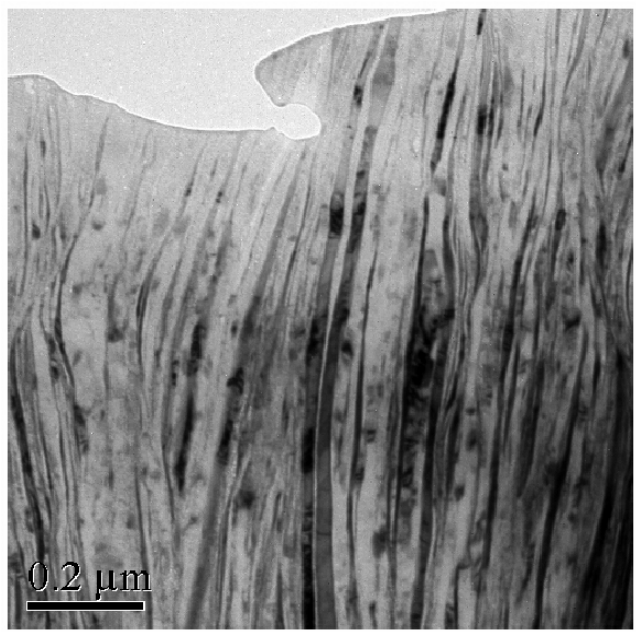

**Figure 1: TEM bright field image showing the cross sectional microstructure of the cold rolled Al/Ni multilayer.**

During the cold rolling process, both Al and Ni foils are severely elongated in the rolling plane, resulting in a multilayer system with layer thickness in the nanometer range as indicated in Fig. 1. The average thickness of the layers is about 10 nm to 50 nm after 50 F&R-cycles. XRD data indicate that only the fcc Al and fcc Ni phases are present in the as-rolled multilayers, indicating further that no detectable phase formation reaction between Al and Ni occurred during the repeated deformation process (Fig. 2).

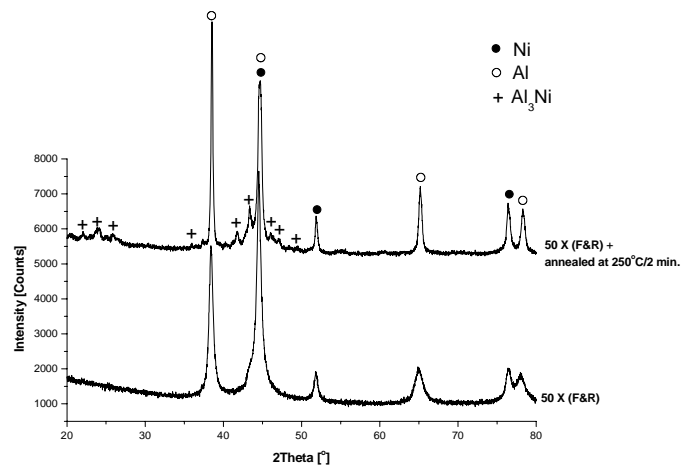

**Figure 2 : X-ray diffraction results of the Al/Ni multilayer in the as-prepared state and after short-time annealing, which causes the formation of the Al$_3$Ni-phase as indicated by the additional Bragg-reflections.**

Some pure Ni and pure Al regions were analysed by 3D-AP (not shown here), but a significant intermixing of Al and Ni was also revealed (Fig. 3). Regions containing between 15 at.% Ni and even up to 60at.% Ni were indeed detected, especially where the layer thickness was in the range of 10 nm or less (measured by field ion microscopy prior to 3D-AP analysis). Since both XRD and selected area electron diffraction in the TEM revealed only fcc Al and fcc Ni in the as prepared sample, these mixed regions are attributed to metastable or – for the highest Ni-content – even non-equilibrium supersaturated solid solutions. It is important to note that no impurity was detected on mass spectra, indicating that there was no contamination during the F&R process. Up to now it has been reported that cold rolled multilayers exhibit sharp interfaces [10-12, 12, 16], e.g. without significant mixing like deposited or sputtered multilayers [19, 20]. However, in these earlier studies of F&R multilayers the local distribution of the constituents was not checked at the atomic scale. On the other hand, the formation of supersaturated solid solutions by ball milling was reported for Al-Ni powders [5, 28], and for many different systems [6, 7]. In the present case, it seems clear from the composition distribution in Fig. 3 that the mixing process is controlled by the diffusion of Ni in Al in agreement with ball milling data [5]. From the composition profile of Fig. 3, the effective diffusion distance R* is estimated as about 10 nm. Then, considering the roll geometry and the roll velocity, the total deformation time is $\tau_d \approx 30$ s. This yields an effective diffusion coefficient $D^* = R^{*2} / \tau_d = 1/3 \; 10^{-18}$ m$^2$s$^{-1}$. Extrapolation of experimental thermal diffusion data shows that similar values are measured at temperatures above 600 K [29]. Such a temperature increase is unrealistic because of the low strain rate during rolling and the high thermal conductivity of rolls. Thus, thermal diffusion does not account for the observed mixing and other mechanisms like forced mixing [3] or deformation induced vacancies [8] may promote the atomic mobility.

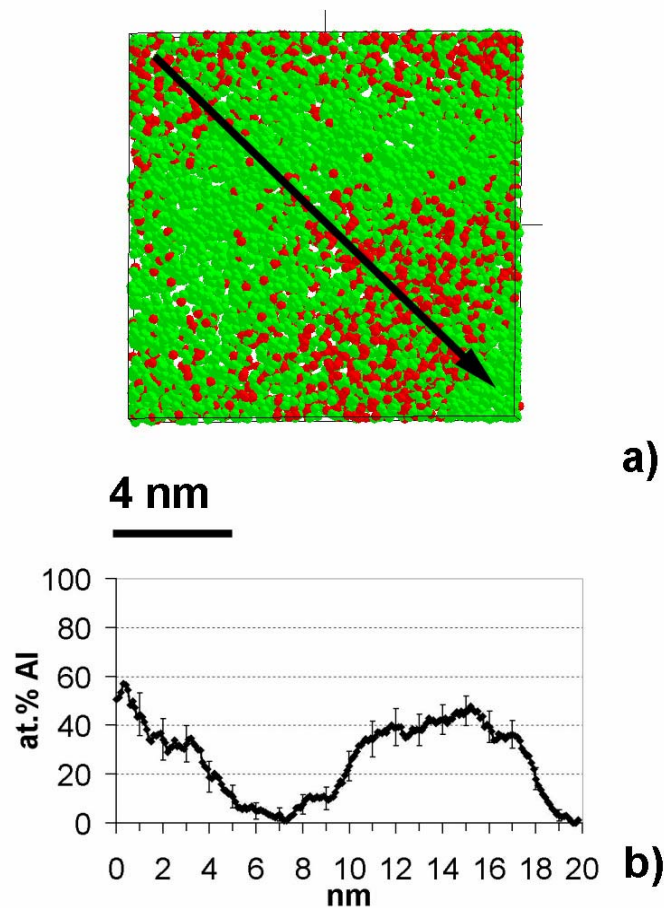

**Figure 3 : (a) 3D reconstruction of an analysed volume in the as prepared Al/Ni multilayer (Ni green, Al red) ; (b) composition profile along the direction indicated by the arrow (sampling box thickness is 1 nm).**

Yet, it is also revealing that intermixing was observed only in regions of the sample where the layer thickness is below 10 nm. This observation indicates that the intermixing is directly coupled with the plastic deformation. The regions with the smallest layer thickness after the repeated rolling deformation are regions of high stress concentration where "necking" [4] occurs. This phenomenon has been reported to occur in Al/Ni multilayer co-deformation and causes enhanced layer thinning in localized regions [10]. The observation also supports the presence of an intermixing mechanism that is not based on thermal diffusion.

The supersaturated solid solutions analysed in the present study are far away from the solubility limits and thus far away from equilibrium [30]. One should also consider that both, mixed regions and concentration gradients up to $10^8$ m$^{-1}$ (Fig. 3(b)) can significantly modify the kinetics of phase formation and – more importantly - can prevent the nucleation of intermediate phases [32]. Thus, to clarify this point, samples have been annealed for a short time at a comparatively low temperature. In accordance with existing reports in the literature [11, 12, 14, 17, 18, 22, 23], the intermetallic Al$_3$Ni phase formed first during the annealing treatment at low temperature, but the XRD data also show that – as intended - the reaction between Al and Ni was by far not completed after annealing for 2 min. at 250°C (Fig. 2). These three phases were also analysed by 3D-AP after annealing: essentially pure Ni regions (99.85±0.04 at.% Ni), regions with a composition that corresponds to the Al$_3$Ni phase (25.5±0.35 at.% Ni) and almost pure Al regions (Ni concentrations in a range of 1 to 2 at.%). Together with the results from XRD we can conclude that the regions with a composition close to 25 at. % Ni actually correspond to the intermetallic Al$_3$Ni phase. As expected form the fine-scaled layer structure observed by TEM (Fig. 1), some interfaces between Al and Al$_3$Ni were intercepted with 3D reconstructed volumes analysed by the atom probe (Fig. 4(a)) and data analysis reveals that concentration profiles across such interfaces are extremely sharp (Fig. 4(b)).

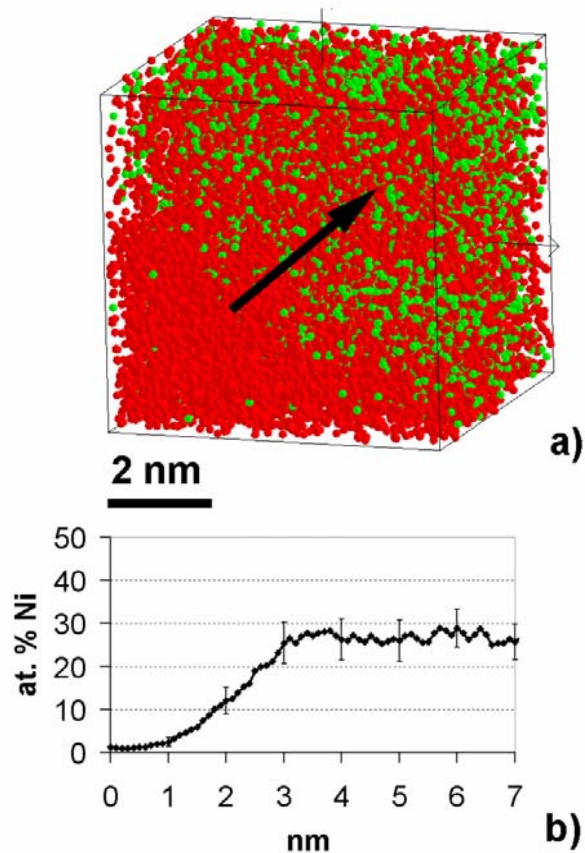

**Figure 4 :** a) 3D reconstruction of an analysed volume in the Al/Ni multilayer annealed at 250°C for 2 min (Ni green, Al red) ; (b) composition profile across the Al / Al$_3$Ni interface along the direction indicated by the arrow (sampling box thickness is 1 nm).

The observed nucleation of the Al$_3$Ni phase during short time annealing is consistent with published data showing that this phase is the first to grow in annealed Al/Ni multilayers [11, 12, 14, 17, 18, 22, 23]. It is however recognised that intermixing is a necessary initial process that needs to occur before any phase formation can proceed [17, 21]. In cold rolled multilayers, this first step is not thermally activated but induced by the severe plastic deformation prior to annealing. This feature might explain why we measured an onset temperature of 468 K (DSC, 20K/min) for the Al$_3$Ni phase formation, while for deposited multilayers, reported data obtained at a similar heating rate are higher (513K for a period of 11.42 nm [27] and 476K for a period of 10 nm [16]). Thus, the present results indicate that the nucleation of Al$_3$Ni is kinetically favoured in Al-Ni mixed regions. However, limited preliminary redistribution of solute is still necessary, since mechanically mixed zones mostly do not have the Al$_3$Ni stoichiometry. It is important to note that the Al$_3$Ni-phase forms first, although the compositions of the intermixed regions are different in different parts of the sample and are partly even close to $Al_{50}$-$Ni_{50}$, i.e. the composition of the B2 ordered AlNi-phase. Moreover, it is important to note that the same phase formation sequence was observed upon annealing as in most of the reports in the literature – keeping in mind that the sample processed by F&R has been intermixed at ambient conditions before the annealing treatment. Thus, the present results confirm that the $Al_3Ni$-phase is kinetically favoured and that the competition in nucleation kinetics determines the evolution of the first phase to form from nanosize interface regions to microscopically observable volumes. These results also indicate the stability of the outcome of the kinetic competition in Al-Ni, indicating further that compositional variation might offer the only way to modify the phase formation sequence.

In conclusion, chemical analyses at the atomic scale of cold rolled Al/Ni multilayers clearly revealed significant intermixing induced by severe plastic deformation. This non-thermal diffusion leads to the formation of metastable or non-equilibrium solid solutions with various compositions. The analysis of the early stage of intermetallic phase formation shows that even if some atomic redistribution is needed prior to the nucleation of $Al_3Ni$, the reaction is faster than in deposited multilayers where interfaces are sharp. The present work confirms that the $Al_3Ni$ phase is strongly favoured by nucleation kinetics so that even the presence of mixed regions with largely different compositions does not result in the initial formation of a different intermetallic phase.

**Acknowledgement**

Dr. H. Roesner (*Institute of Nanotechnology, Forschungszentrum Karlsruhe, Germany*) is gratefully acknowledged for the TEM investigation of the Al/Ni multilayers.

**References**

[1]  K.N. Tu, U. Gösele, Mat. Res, Soc. Symp. Proc. 19 (1983) 375.

[2]  P.J. Desré, A.R. Yavari, Phys. Rev. Lett. 64 (1990) 1533.

[3]  P. Bellon, R.S. Averback, Phys. Rev. Lett. 74 (1995) 1819.

[4]  F. Bordeaux, R. Yavari, Z. Metallkde. 81 (1990) 130.

[5]  F. Cardellini, G. Mazzone, A. Montone and M. Vittori Antisari, Acta Metall Mater 42 (1994) 2445.

[6]  E. Ma, M. Atzmon, Materials Chemistry and Physics 39 (1995) 249.

[7]  C. Suryanarayana, Prog. Mat. Sci. 46 (2001) 1.

[8]  X. Sauvage, F. Wetscher and P. Pareige, Acta Mater 53 (2005) 2127.

[9]  G. Wilde, H. Sieber, J.H. Perepezko, Scripta Mater. 40 (1999) 779.

[10] H. Sieber, J.H. Perepezko, Journal of Material Science Letters 18 (1999) 1449.

[11] H. Y. Kim, D. S. Chung, S.H. Hong, Materials Science and Engineering A 396 (2005) 376.

[12] H. Sieber, J.S. Park, J. Weissmueller and J.H. Perepezko, Acta Materialia 49 (2001) 1139.

[13] E.G. Colgan, Mater Sci Rep 5 (1990) 44.

[14] L. Battezzati, C. Antonione, F. Fracchia, Intermetallics 6 (1995) 67.

[15] H. Y. Kim, D. S. Chung, S.H. Hong, Scripta Materialia 54 (2006) 1715.

[16] E. Ma, C.V. Thompson, L.A. Clavenger, K.N. Tu, Applied Physic Letter 57 (1990) 1262.

[17] M.H. Da Silva Bassani, J.H. Perepezko, A.S. Edelstein, R.K. Everett, Scripta Mater 37 (1997) 227.

[18] L. Battezzati, P. Pappalepore, F. Durbiano, I. Gallino, Acta Mater 47 (1999) 1901.

[19] T. Jeske, G. Schmitz, Mater. Sci. Eng. A327 (2002) 101.

[20] T. Jeske, M. Seibt, G. Schmitz, Mater. Sci. Eng. A353 (2003) 105.

[21] E. Ma, C.V. Thompson, L.A. Clevenger, J. Appl. Phys. 69 (1991) 221.

[22] E.G. Colgan, J.W. Mayer, Nucl. Instr. Methods B17 (1986) 242.

[23] E. Ma, M.-A. Nicolet, M. Nathan, J. Appl. Phys. 65 (1989) 2703.

[24] C. Michaelsen, G. Lucadamo, K. Barmak, J. Appl. Phys. 80 (1996) 6689.

[25] M. Atzmon, Phys. Rev. Lett. 64 (1990) 487.

[26] F. Cardellini, G. Mazzone, A.V. Antisari, Acta Mater. 44 (1996) 1511.

[27] A.S. Edelstein, R.K. Everett, G.R. Richardson, S.B. Qadri, J.C. Foley, J.H. Perepezko,

Mat. Sci. Eng. A 195 (1995) 13.

[28] J.S.C Jang, C.C Koch, J. Mat. Res. 5 (1996) 498.

[29] G. Erdelyi, D.L. Beke, F.J. Kedves, I. Gödeny, Phil. Mag. B 38 (1978) 445.

[30] T.B. Massalski, Binary Alloy Phase Diagrams, American Society for Metals, Metals Park, OH, 1986, 142.

[31] A.R. Yavari, O. Drbohlav, Mat. Trans. JIM 36 (1995) 896.